\documentclass[a4paper]{jpconf}

\usepackage{amssymb,amsmath,verbatim,color,mathtools,enumitem,xcolor,hyperref,graphicx}
\usepackage{hyperref}
\definecolor{linkcolor}{rgb}{0.0,0.3,0.5}
\hypersetup{colorlinks=true,linkcolor=linkcolor,citecolor=linkcolor,filecolor=linkcolor,urlcolor=linkcolor}

\begin{document}
\title{Surprises from the spins: astrophysics and relativity with detections of spinning black-hole mergers}

\author{Davide Gerosa\footnote{Einstein Fellow}}

\address{TAPIR 350-17, California Institute of Technology,\\
1200 E California Boulevard, Pasadena, CA 91125, USA}

\ead{dgerosa@caltech.edu}

\begin{abstract}
Measurements of black-hole spins are of crucial importance to fulfill the promise of gravitational-wave astronomy. On the astrophysics side, spins are perhaps the cleanest indicator of black-hole evolutionary processes, thus providing a preferred way to discriminate how LIGO's black holes form. On the relativity side, spins are responsible for peculiar dynamical phenomena (from precessional modulations in the long inspiral to gravitational-wave recoils at merger) which encode precious information on the underlying astrophysical processes. I present some examples to explore this deep and fascinating interplay between spin dynamics (relativity) and environmental effects (astrophysics). Black-hole spins indeed hide remarkable surprises on both fronts: morphologies, resonances, constraints on supernova kicks, multiple merger generations and more...
\vspace{0.1cm}\\
These findings were presented at $12^{\rm th}$ Edoardo Amaldi Conference on Gravitational Waves, held on 
July 9-14, 2017 in Pasadena, CA, USA. 
\end{abstract}

\section{Introduction}

The second-generation gravitational-wave (GW) detectors Advanced LIGO and Virgo are taking us into the age of observational strong-field gravity. The detections of black-hole (BH) binaries has now become routine with 4.87 events observed during Advanced LIGO/Virgo first and second observing runs (4 confirmed detections \cite{2016PhRvL.116f1102A,2016PhRvL.116x1103A,2017PhRvL.118v1101A,2017arXiv170909660T} and 1 candidates with 87\% probability of being of astrophysical origin \cite{2016PhRvX...6d1015A}). These observations are giving us unique insights on a sector of the Universe we hardly knew existed. Three of the detected binaries (GW150914, GW170104 and GW170814 \cite{2016PhRvL.116f1102A,2017PhRvL.118v1101A,2017arXiv170909660T} have component masses $\sim 30 M_\odot$, largest than all stellar-mass BHs known from X-ray binary observations \cite{2010ApJ...725.1918O}. The event GW151226  \cite{2016PhRvL.116x1103A} has a lower mass, but a confirmed detection of BH spin. 

Despite the low statistics, these first BH detections are already providing invaluable information on the life of their stellar progenitors. For instance, the unexpected high mass of some of the events can only be modeled if systems form in a low metallicity environment ($Z\lesssim 0.5 Z_\odot$), where stellar winds are suppressed \cite{2016ApJ...818L..22A,2016MNRAS.463L..31L}. More observations will become available in the next few years as detectors keep on being improved, thus allowing us to enter the large-statistics regime and measure details of the underlying astrophysical populations of BHs in binaries.

\begin{figure}
\includegraphics[width=18pc]{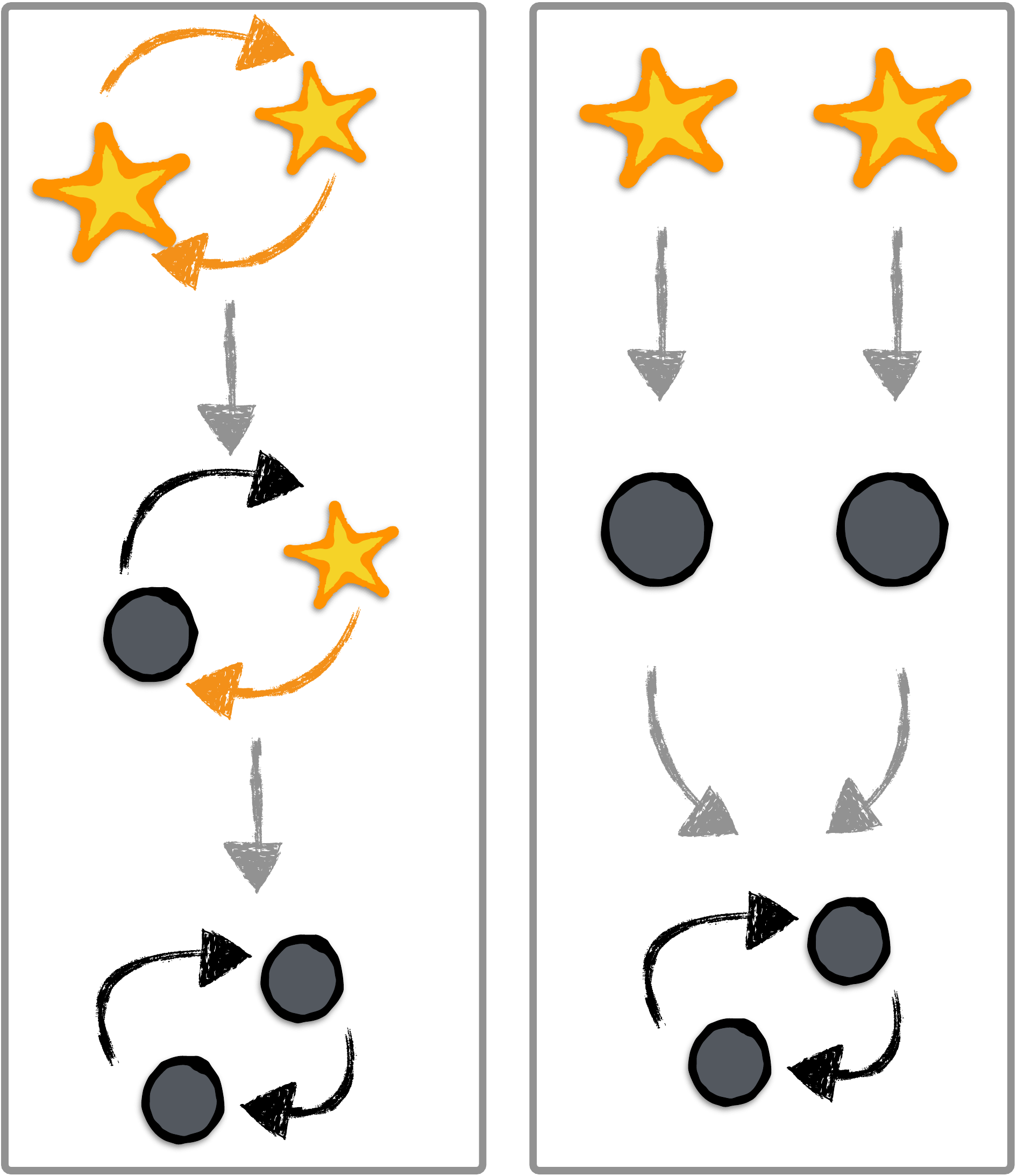}\hspace{2pc}%
\begin{minipage}[b]{12.5pc}\caption{\label{diagram}Two main classes of stellar-based BH formation models. BH binaries can form \emph{in the field} from binary stars (left): both stars collapse and form BHs without unbinding the binary, leaving a GW source behind. BH in binaries can also form separately, from two uncorrelated stars, and later meet because of dynamical interactions (right). The latter class of models typically requires a dense stellar environment to efficiently assemble GW sources.\vspace{0.05cm}}
\end{minipage}
\end{figure}

There are two main classes of stellar-based BH-binary formation models, depending on whether the two BHs have spent all their lives together as a binary star \cite{2014LRR....17....3P}, or they rather evolved separately and later met to form a BH binary \cite{2013LRR....16....4B} (see Fig.~\ref{diagram} for a schematic diagram). In the first scenario (Fig.~\ref{diagram}, left panel), two massive stars forming a binary system \emph{in the field} both evolve into BHs and leave a compact binary behind. There are several key steps involved in this class of models, notably a common envelope phase and natal kicks imparted to the BHs at birth. In the second class of models (Fig.~\ref{diagram}, right panel), a dense stellar environment such as a globular or open cluster is invoked to assemble BHs through dynamical interactions. 

We face a pressing question in GW astrophysics: can we distinguish between these formation channels using current and future BH binary observations? Can we infer some of the key parameters entering these models? In other terms: \emph{do BH binaries remember how they form?}

The BH mass distributions predicted by several formation models tend to overlap, meaning that $\emph{O}(100)$ observations will be necessary before convincing constraints can be extracted using mass measurements only \cite{2015ApJ...810...58S,2017arXiv170407379Z}. Eccentricity and redshift measurements could provide strong constraints for specific formation models (cf. \cite{2012ApJ...757...27A} and \cite{2017MNRAS.471.4702B}, respectively), but are perhaps not going to be very informative for the most favorable scenarios. Spins, on the other hand, are arguably the cleanest observables to distinguish BH formation channels.
If the two BHs were together as a binary star, the relative orientations of their spins and the binary's orbital angular momentum are expected to reflect processes able to (mis)align the stellar spins (such as BH natal kicks and tidal interactions  \cite{2000ApJ...541..319K,1981A&A....99..126H,2013PhRvD..87j4028G}). Conversely, if BHs form in clusters, their spin orientation is expected to be uncorrelated and isotropically distributed.

Here I present some first attempts to exploit the spin dynamics to extract astrophysical information from current and future BH binary observations. Sec.~\ref{twospins} summarizes the phenomenology of two-spin BH binaries and introduces the spin morphology as a tool to directly infer the spin orientations at BH formation from those observed by LIGO/Virgo at $\sim 100$ Hz. Sec.~\ref{remember} briefly describes how spins can be used to extract information on BH natal kicks and the occurrence of multiple merger generations. 
We draw our conclusions in Sec.~\ref{conclusions}. Unless otherwise stated, we use geometric units $c=G=1$.

\section{Timescales and morphologies}
\label{twospins}



The phenomenology of spinning BH binaries is better understood by separating the various timescales of the problem. The two BHs orbit about each other on a timescale $t_{\rm orb}\propto (r/M)^{3/2}$, the spins and the orbital angular momentum precess on a timescale $t_{\rm pre}\propto (r/M)^{5/2}$, while the orbit shrinks on $t_{\rm rad}\propto (r/M)^{4}$ \cite{1994PhRvD..49.6274A} (here $r$ is the binary separation and $M=m_1+m_2$ is the total mass). In the post-Newtonian (PN) regime $r/M\gg 1$ and therefore $t_{\rm orb}\ll t_{\rm pre} \ll t_{\rm rad}$. The first inequality  $t_{\rm orb}\ll  t_{\rm pre}$ has been used since the very early developments of the PN formalism,
 leading to the orbit-averaged formulation of the PN equations of motion. In general, the relative orientations of the spins $\mathbf{S_i}= \chi_i m_i^2 \mathbf{\hat S_i}$  (with $i=1,2$) and the binary's angular momentum $\mathbf{L}=m_1 m_2 \sqrt{r/M} \mathbf{\hat L}$ depend on three variables, which can be chosen the be the angles $\theta_i$ between $\mathbf{S_i}$ and $\mathbf{L}$, and the angle $\Delta\Phi$ between $\mathbf{\hat S_1}\times \mathbf{\hat L}$ and $\mathbf{\hat S_2}\times \mathbf{\hat L}$ (c.f. Fig. 1 in \cite{2015PhRvD..92f4016G}). Only recently \cite{2015PhRvD..92f4016G,2015PhRvL.114h1103K}, we presented an analysis of the BH binary dynamics which fully exploits the second inequality  $t_{\rm pre} \ll t_{\rm rad}$, and allows us to formulate the equations in a \emph{precession-averaged} fashion. This was made possible by the presence of two additional constants of motion: (i) the effective spin $\chi_{\rm eff}$   (or $\xi$ in the notation of \cite{2015PhRvD..92f4016G,2015PhRvL.114h1103K}) is constant on both  $t_{\rm pre}$ and $t_{\rm rad}$ \cite{2008PhRvD..78d4021R}, and (ii) the magnitude $J=|\mathbf{L}+\mathbf{S_1}+\mathbf{S_2}|$ is 
 constant on $t_{\rm pre}$ and slowly evolves on $t_{\rm rad}$. On the precession time, the entire dynamics can be described in terms of the single variable $S=|\mathbf{S_1}+\mathbf{S_2}|$. 
 
The equations of motion can now be solved with straightforward numerical integrations \cite{2015PhRvD..92f4016G}, or even analytically \cite{2017PhRvL.118e1101C}. The precession dynamics can be classified in terms of the qualitative evolution of the spin angles $\theta_1$, $\theta_2$ and $\Delta\Phi$. While $\theta_1$ and $\theta_2$ always evolve monotonically during each precession cycle, three different \emph{morphologies} can be found in $\Delta\Phi$. Orbits of $\Delta\Phi$ can circulate in the whole range $[0,\pi]$, librate about $0$ (never reaching $\pi$) or librate about $\pi$ (never reaching $0$). The crucial point here is that, while the spin angles $\theta_1$, $\theta_2$ and $\Delta\Phi$ vary on the short timescale $t_{\rm pre}$, the spin morphology does not. It is only set by the values of $\chi_1$, $\chi_2$, $q=m_2/m_1$, $\chi_{\rm eff}$ and $J$, and is therefore constant on $t_{\rm pre}$. Secular variations of $J$ on $t_{\rm rad}$ might cause transitions between the three morphologies, generically acting towards bringing librating binaries towards the two circulating morphologies.

Quantities subject to spin precession are typically quoted at some reference frequency in GW searches (set to $f_{\rm ref}=20$ Hz in current LIGO analyses). We argue one should always try to estimate quantities varying on the longest timescales of the problem, so that the values reported better characterize the observed systems. From this point of view, the effective spin parameter $\chi_{\rm eff}$ is an excellent choice, because it is constant on both $t_{\rm pre}$ and $t_{\rm rad}$ to high PN order. The spin morphology is also an interesting quantity, because it only varies on $t_{\rm rad}$. On the contrary, the spin angles vary on the shorter timescale $t_{\rm pre}$.

\begin{figure}[t]
\centering
\includegraphics[width=11pc]{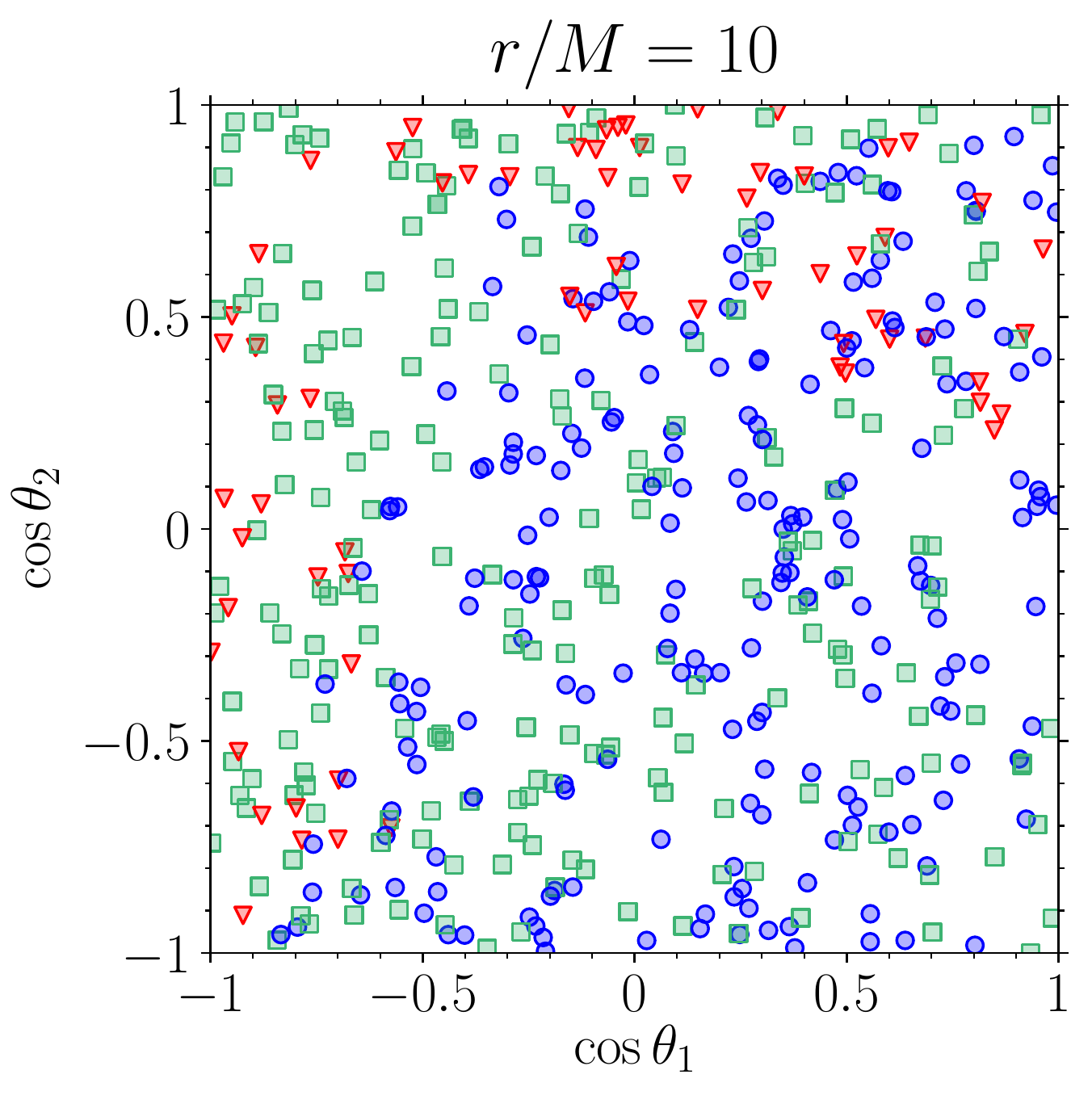}$\;\;$
\includegraphics[width=11pc]{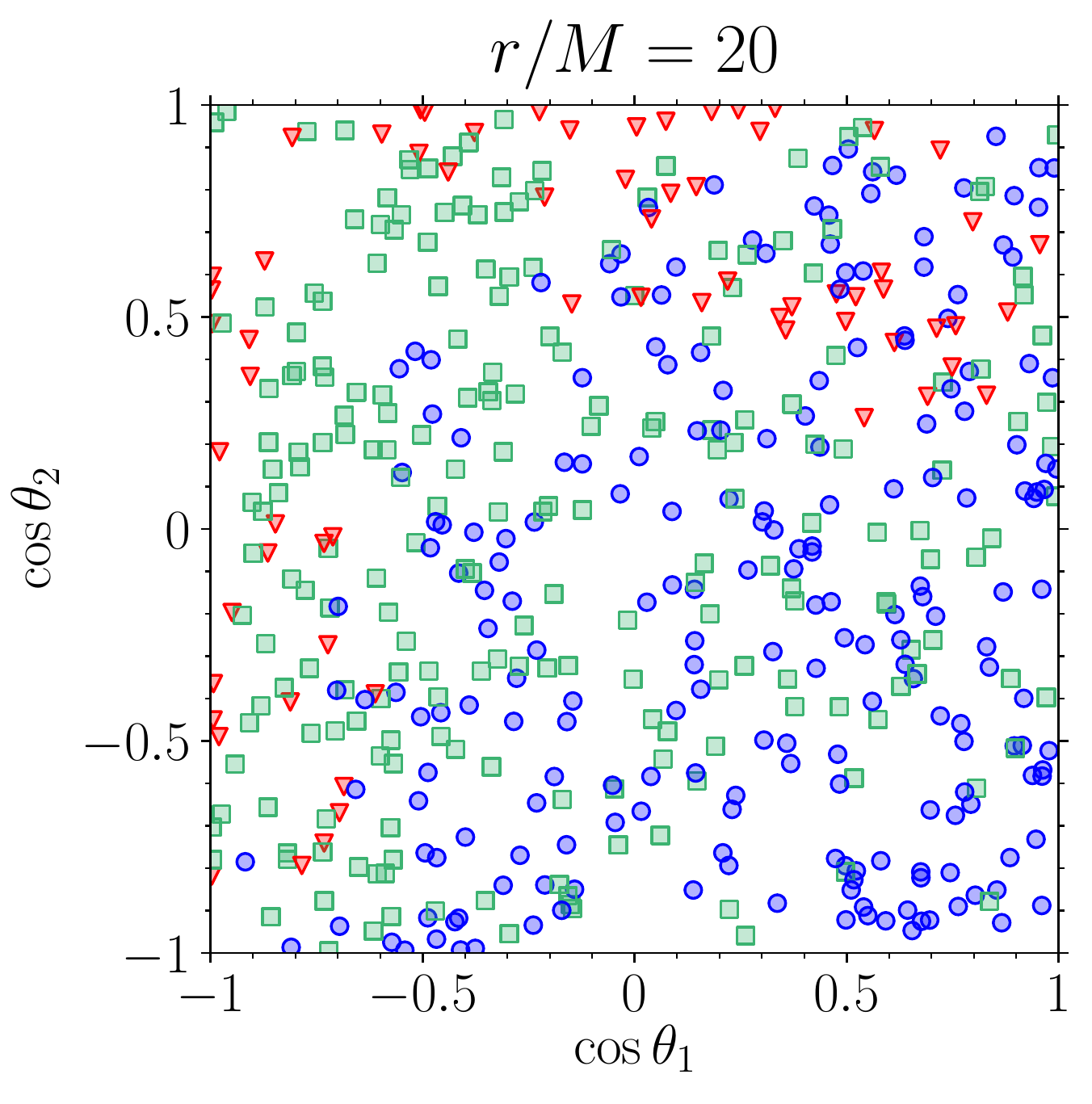}$\;\;$
\includegraphics[width=11pc]{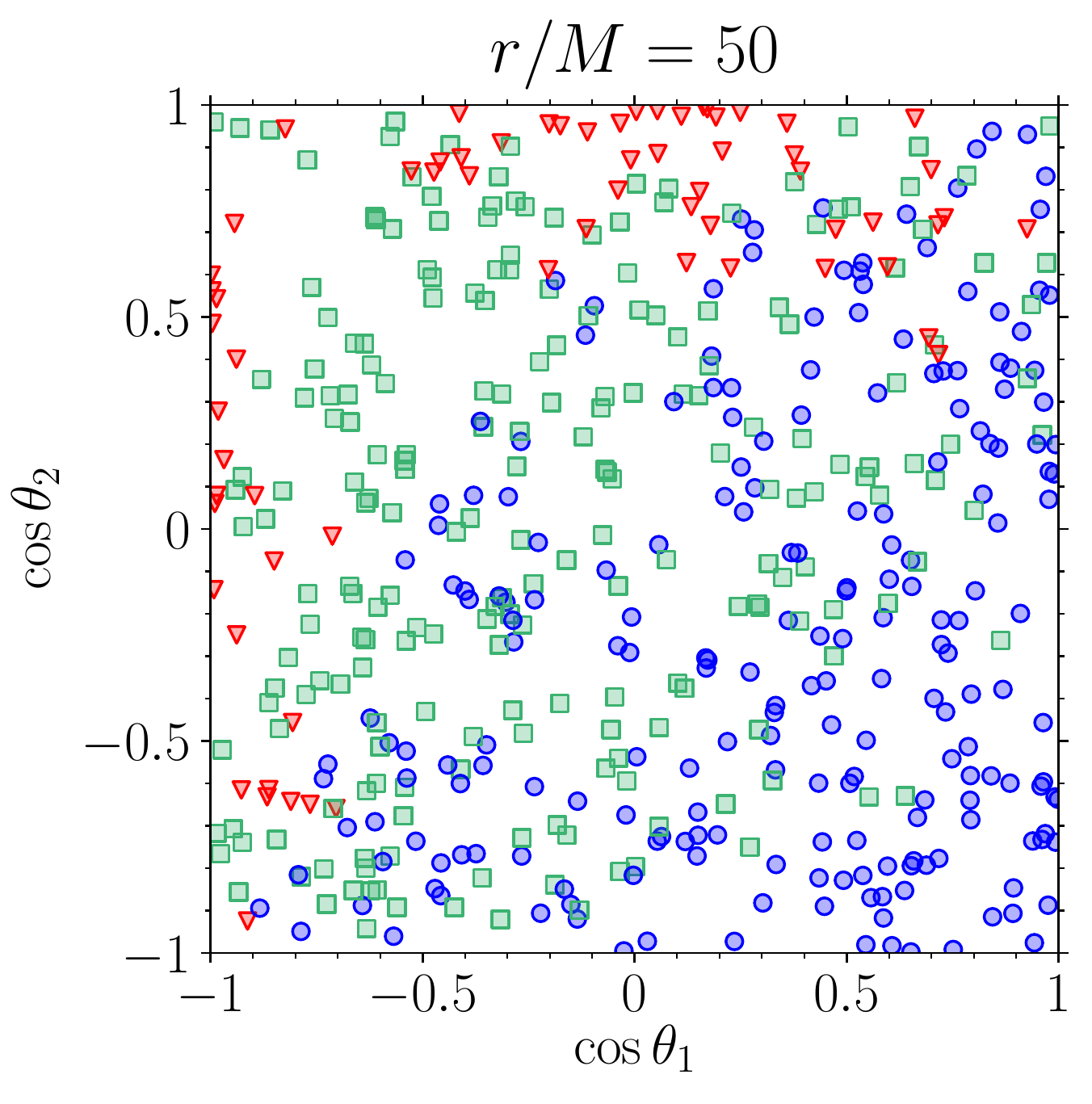}\\
\vspace{0.1cm}
\includegraphics[width=11pc]{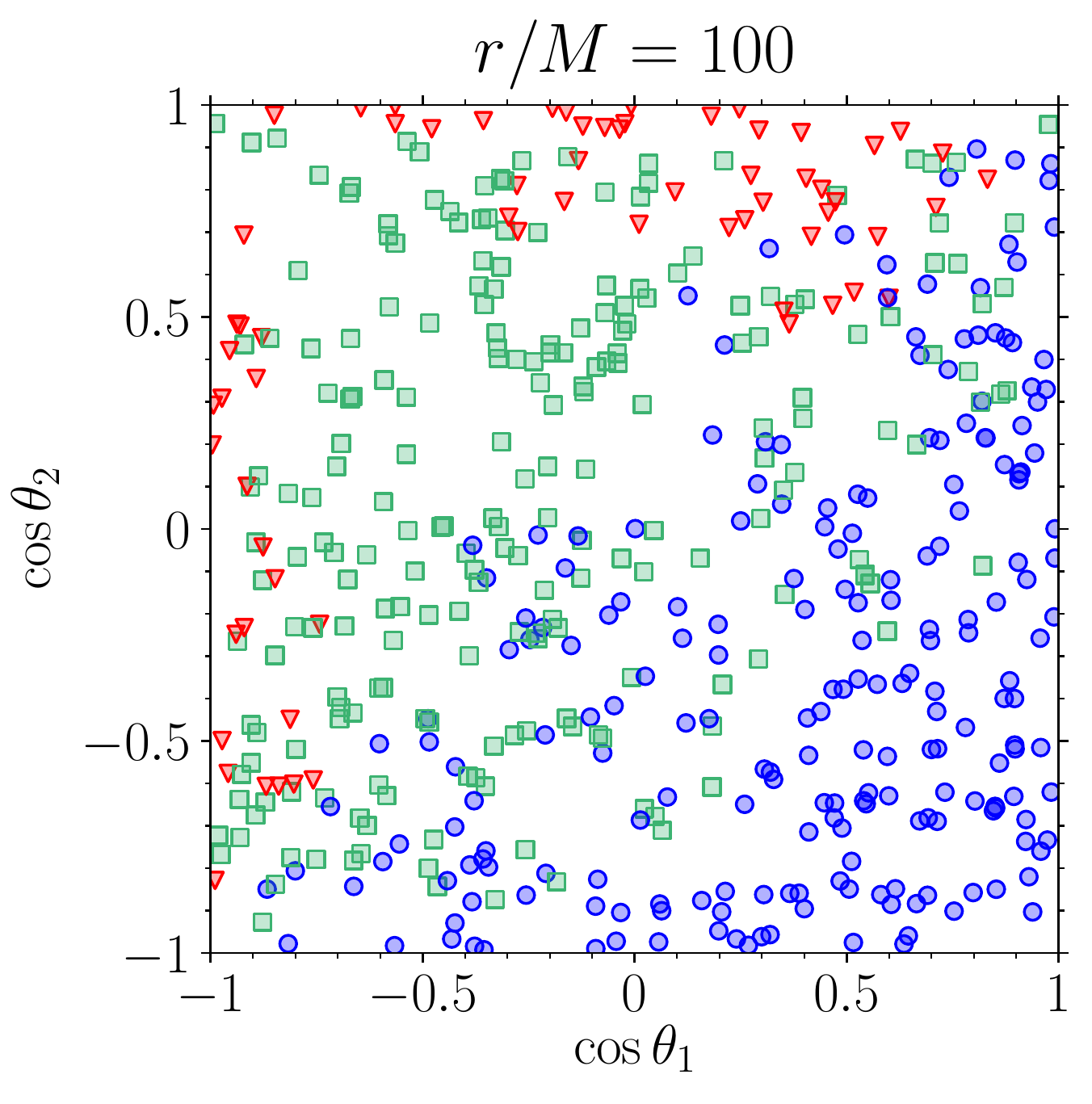}$\;\;$
\includegraphics[width=11pc]{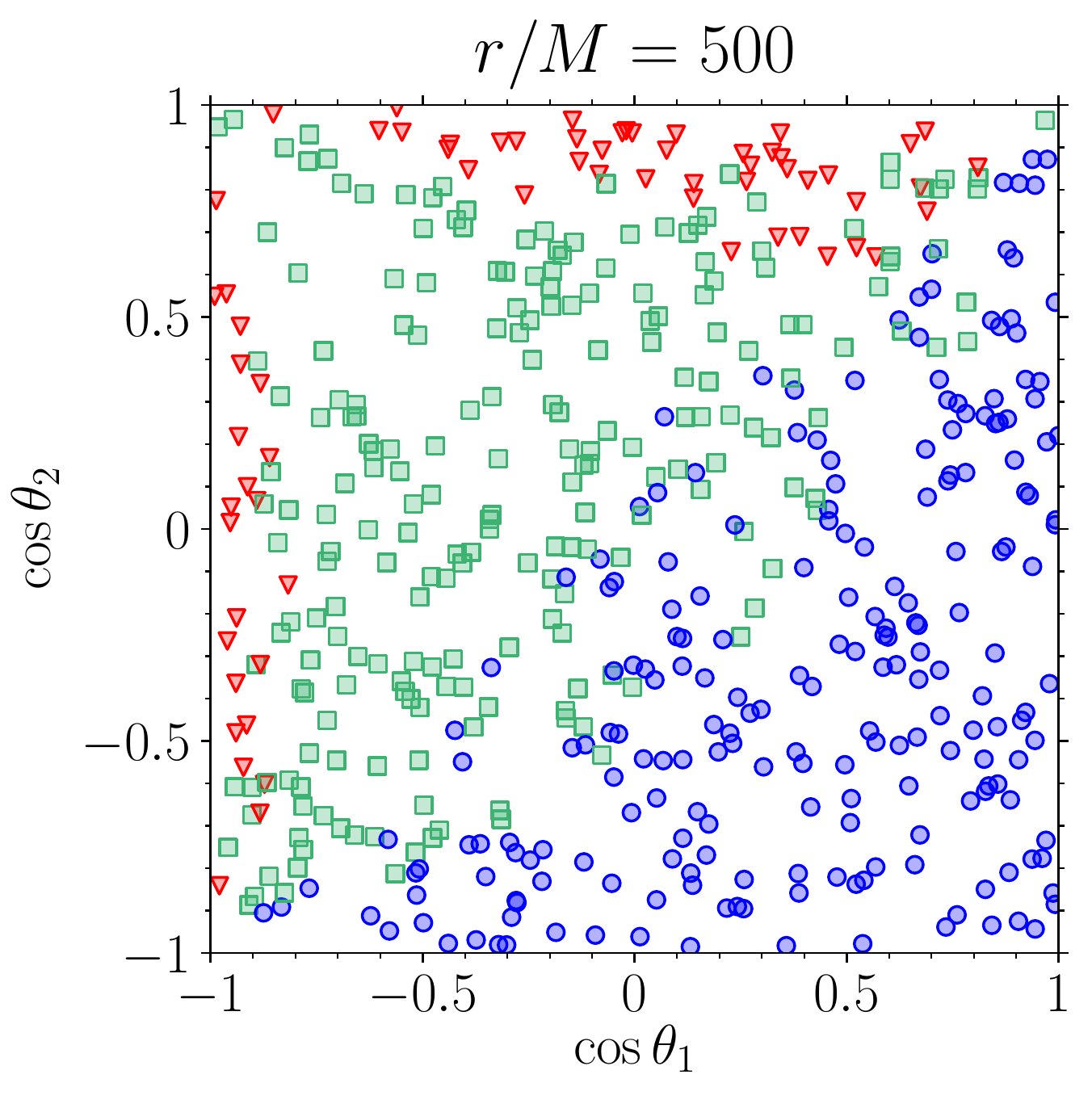}$\;\;$
\includegraphics[width=11pc]{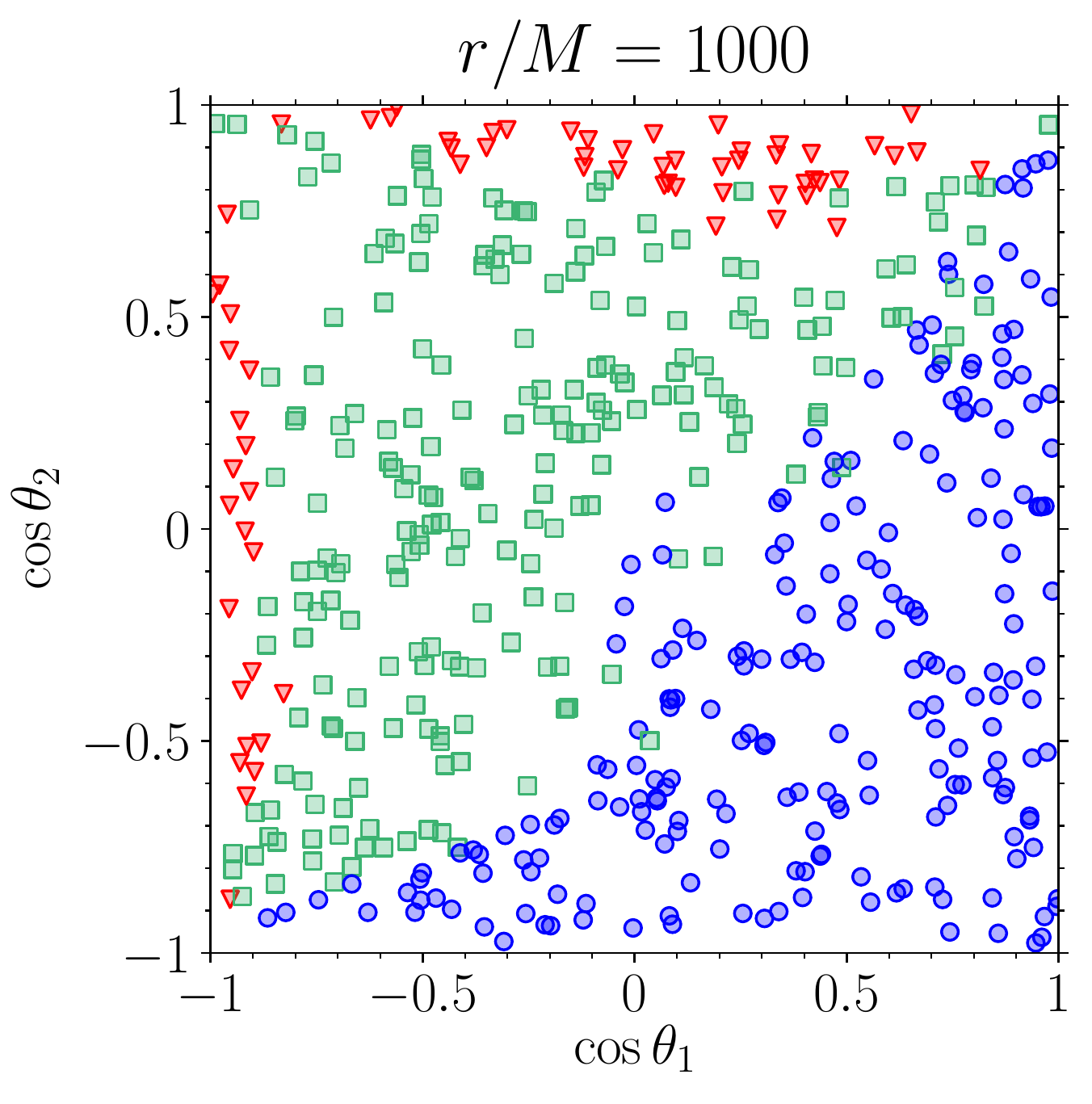}\\
\vspace{0.1cm}
\includegraphics[width=11pc]{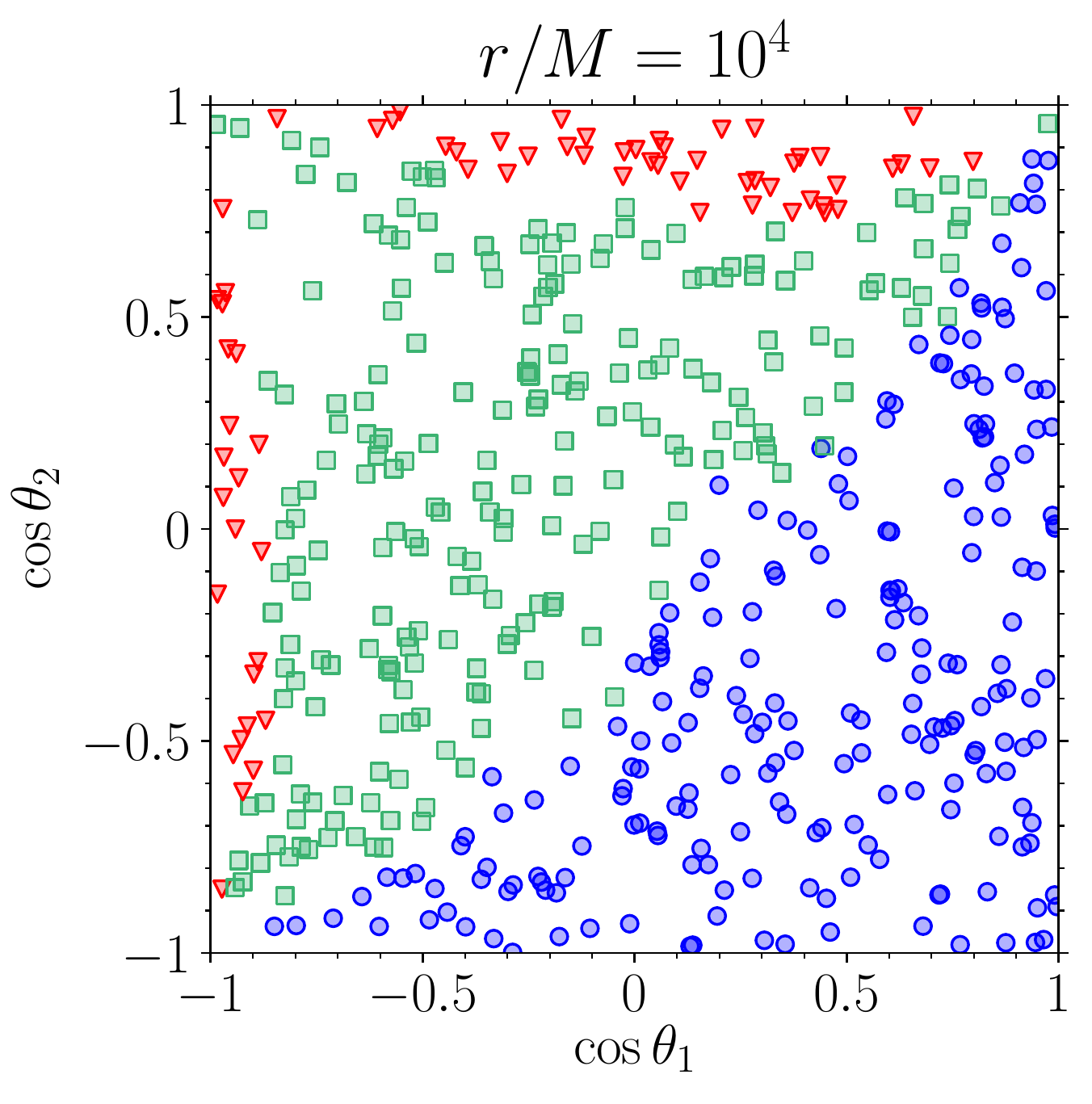}$\;\;$
\includegraphics[width=11pc]{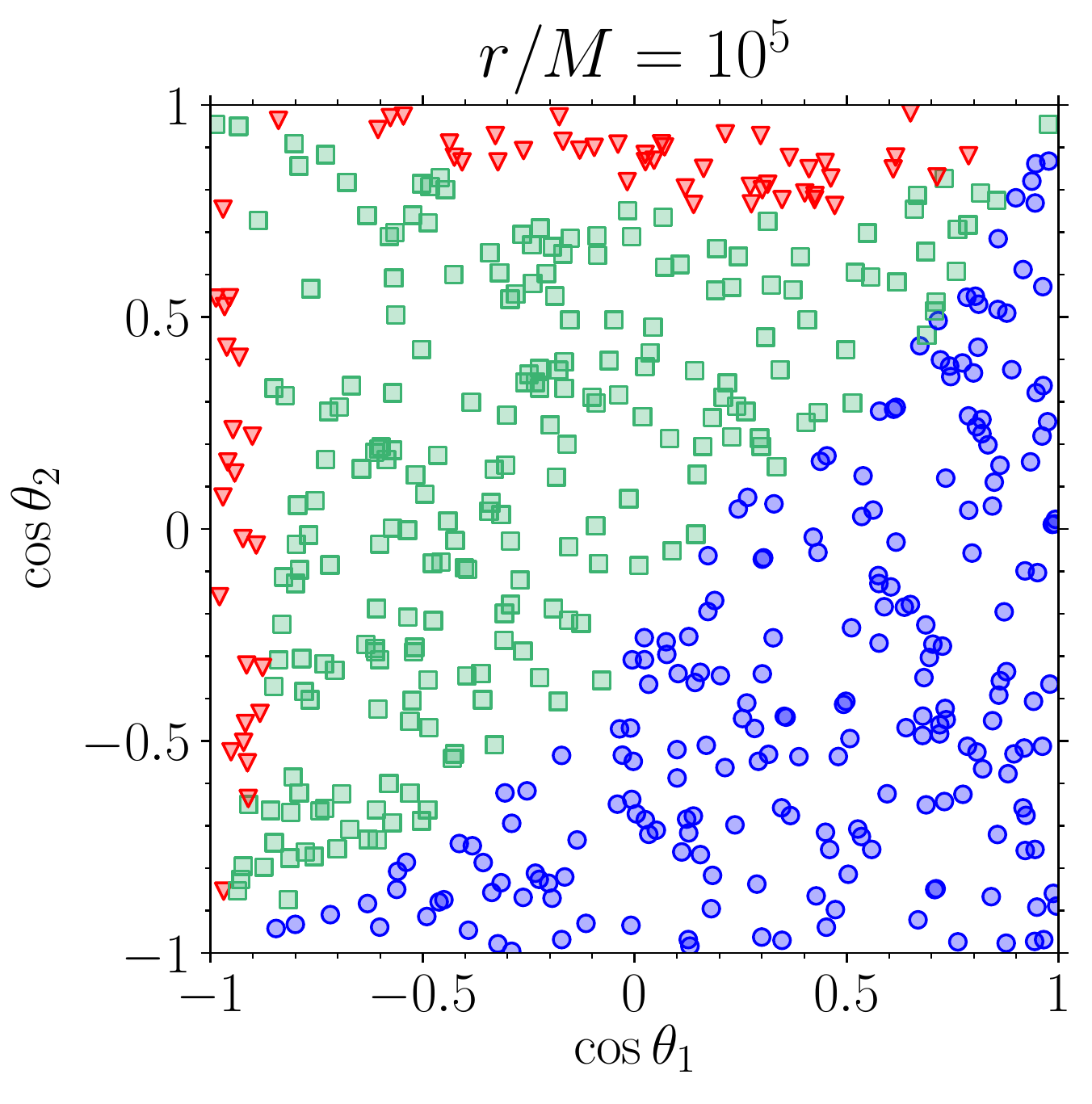}$\;\;$
\includegraphics[width=11pc]{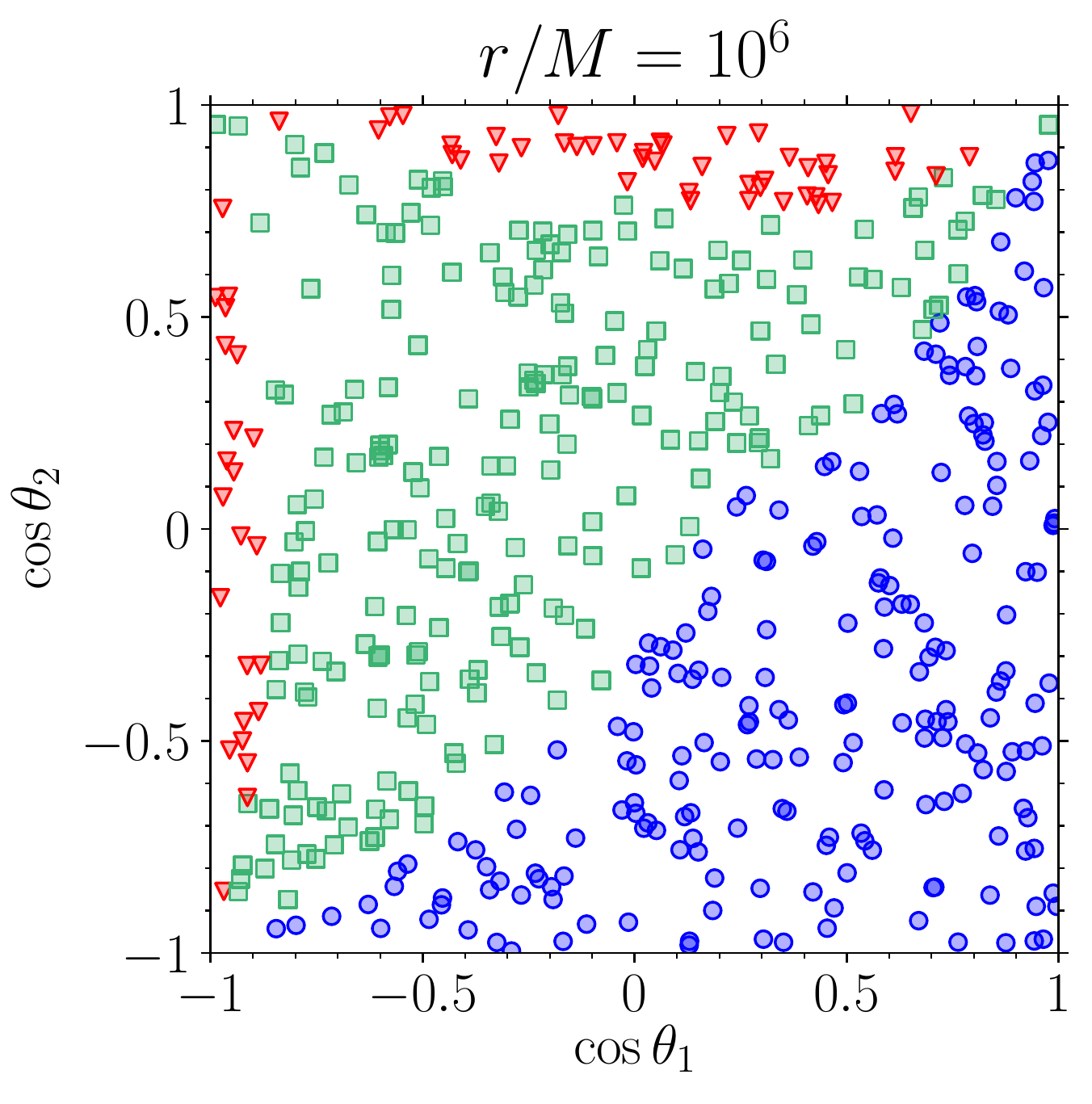}\\
\caption{\label{evolution}
Precession-averaged PN evolutions of BH binaries. We evolve 500 systems with $\chi_1=\chi_2=1$ and $q=0.8$ with isotropic spin directions at large separations. The evolution is shown backwards, from $r=10M$ (top-left panel) to $r=10^6M$ (bottom-right panel), in the plane defined by the two spin orientations $\cos\theta_i=\mathbf{\hat S_i} \cdot \mathbf{\hat L}$. Each binary is colored according to its spin morphology  at $r=10M$: blue for binaries librating about $\Delta\Phi=0$, green for binaries circulating in $\Delta\Phi\in[0,\pi]$ and red for binaries librating about $\Delta\Phi=\pi$. Binaries with a given morphology at small separation (detection) originate from precise regions in the ($\theta_1$,$\theta_2$) plane at large separation (formation). This figure was produced using the public python module  \textsc{precession} \cite{2016PhRvD..93l4066G}. An animated version is available at \href{http://www.davidegerosa.com/spinprecession}{www.davidegerosa.com/spinprecession}.
}
\end{figure}

\emph{The spin morphology, therefore, encodes information on the precessional dynamics of the binary, but does not vary on the precessional time.} Moreover, if the spin morphology is estimated at, say, $20$ Hz, (well within the detector sensitivity window) it can easily be used to infer information on the spin orientations at BH formation and thus constrain astrophysical models. This point is illustrated in Fig.~\ref{evolution}. We evolved a sample of 500 BH binaries with $\chi_1=\chi_2=1$ and $q=0.8$ from $r=10^6 M$ to $r=10M$. 
Each binary is color-coded according to its spin morphology measured at $r=10M$.
 The spin directions $\theta_i$ at small separations carry little information on the initial configuration of the binary. Conversely, the spin morphology directly tracks the spin orientation at early times: binaries of different colors cluster in different and well-defined regions of the $(\theta_1,\theta_2)$ plane. By measuring the spin morphology in the LIGO band, we will be able to robustly reconstruct information on the spin configuration when the BHs formed.

This idea can be used to constrain specific formation mechanisms that preferentially populate some regions of the $(\theta_1,\theta_2)$ parameter space. An early exploration was presented in \cite{2013PhRvD..87j4028G}, where we showed that stellar-mass BH binaries formed in the field tend to belong to different morphologies if they (do not) undergo mass transfer and tidal interactions during their lives as massive stars.


\section{Spin constraints on black-hole natal kicks and multiple generations}
\label{remember}

Although weakly informative, current LIGO measurements of BH spins already provide some interesting indications. In particular, GW150914, GW170104, GW170814 and LVT151012 are all consistent with zero spins \cite{2016PhRvL.116f1102A,2017PhRvL.118v1101A,2017arXiv170909660T}. The opposite is true for GW151226, where at least one of the two BHs must have been spinning before merger \cite{2016PhRvL.116x1103A,2017arXiv170704637V} . These few measurements can be used to place (model-dependent) constraints on the astrophysics of BH binaries. Here we summarize two of our recent studies on the topic.

\begin{itemize}

\item The ``boxing day" event, GW151226, can be used to infer the magnitude of the recoil velocities that BHs are expected to receive at birth \cite{2017PhRvL.119a1101O}.  At core collapse and BH formation, asymmetric mass and neutrino emission might impart a recoil to the newly-formed compact object. While natal kicks imparted to neutron stars are well constrained by pulsar observations \cite{2005MNRAS.360..974H}, their occurrence on BHs is still uncertain \cite{2012MNRAS.425.2799R,2010MNRAS.401.1514M,2012ARNPS..62..407J}. The primary BH of GW151226 likely had non-zero spin, modestly misaligned with the angular momentum ($25^\circ \lesssim \theta_1 \lesssim 80^\circ$) \cite{2016PhRvL.116x1103A,2017PhRvL.119a1101O}. In a field formation scenario where stellar spins are initially aligned to the orbital angular momentum, this spin misalignment directly tracks the orbital-plane tilt caused by the first recoil \cite{2000ApJ...541..319K,2013PhRvD..87j4028G,2017PhRvL.119a1101O}. We found that a kick $v_k\sim v$ (where $v$ is the orbital velocity at BH formation) must be imparted in order to produce a misalignment of $25^\circ \lesssim \theta_1 \lesssim 80^\circ$, consistent with that observed for GW151226. This constraint can be converted into physical units by averaging over a stellar population \cite{2012Sci...337..444S,2016Natur.534..512B}. We find only values $v_k\gtrsim 45$ km/s can reproduce the observed BH misalignment in $>5\%$ of the realizations. Our analysis was later refined in \cite{2017arXiv170901943W}, where more elaborate, but still model-dependent, constraints are presented using masses \emph{and} spin information of all the events available at the time.

\item One possibility is that the BHs we have been observing do not come directly from the collapse of stars (``first generation''), but rather result from previous BH mergers (``second generation'') \cite{2017PhRvD..95l4046G}. While it is certainly true that some astrophysical interactions are expected to be at play to assemble multiple merger generations (e.g. \cite{2017arXiv170207818M,2017arXiv170901660S}), our approach is agnostic on the specific formation channel. This direction is therefore orthogonal, but complementary, to the usual field vs. cluster debate (Fig.~\ref{diagram}). Once more, the secret to discriminate these models lies in the BH spins. Spins of BHs resulting from a previous merger are mainly set by the angular momentum of the progenitor binary at plunge, with the component spins playing a subdominant role. In practice, this means that the spin distribution of second-generation BHs is expected to be highly peaked at $\chi\sim 0.7$ \cite{2008ApJ...684..822B,2017PhRvD..95l4046G}. 
Multiple merger generations are expected to not only affect the spin distributions, but also BH masses and merger redshifts. On average, second-generation BHs should be more massive, and merge later. In \cite{2017PhRvD..95l4046G}, we combined these insights with Bayesian model-selection techniques to constrain the occurrence of multiple mergers with current and future GW data. We showed that BH binaries from LIGO's first observing run already provide a $1$ to $2\sigma$ constraint favoring first generation mergers. With $20-100$ observations we will be able to constrain these models at $5\sigma$ in $>90\%$ of the realizations.
\end{itemize}

\section{Conclusions}
\label{conclusions}

BH spins are precious observables to infer the astrophysics of BH binaries from GW observations. We showed how information on the BH spins at formation is not erased by the large number of precession cycles that sources undergo before entering the LIGO/Virgo band. Some memory is preserved and encoded into the spin morphologies. The first GW detections are already providing us interesting constraints on the astrophysics of massive stars and BH evolution. We described two examples, where we pioneered the use of GW data to constrain specific formation and evolution mechanisms: (i) the observed misalignment of GW151226 can be used to obtain a (model dependent) constraint on BH natal kicks; and (ii) mass, redshift, and  spin distributions can be used to constrain scenarios where multiple merger generations are responsible for the observed events.

This is just the beginning of a new era in astronomy and relativity. The future is bright, with more events, better detectors, new kind sources, more electromagnetic coincidences and serendipitous discoveries all lying just beyond the horizon.

\section*{Acknowledgments}
These findings were presented in a plenary talk at the $12^{\rm th}$ Edoardo Amaldi Conference on Gravitational Waves, following the awarding ceremony of the 2016 Stefano Braccini Thesis Prize. D.G. would like to thank the Gravitational Wave International Committee and all friends of Stefano Braccini for this award, and encourage future graduates to submit their Theses for consideration.  D.G thanks E.~Berti, M.~Kesden, R.~O'Shaughnessy, U.~Sperhake, and D.~Wysocki, who all contributed to the research presented in these proceedings. D.G. is supported by NASA through Einstein Postdoctoral Fellowship Grant No. PF6-170152 awarded by the Chandra X-ray Center, which is operated by the Smithsonian Astrophysical Observatory for NASA under Contract NAS8-03060.

\section*{References}

\bibliographystyle{iopart-num}
\bibliography{amaldiproc}
\end{document}